\documentclass{ws-p8-50x6-00}

\usepackage{amsmath}

\begin{document}

\title{Dipole-proton $S$-matrix derived from diffractive meson electroproduction}

\author{S. Munier}

\address{INFN, Sezione di Firenze, Sesto Fiorentino (Firenze), 50019, Italy.}



\maketitle

\abstracts{
Through a determination of the $S$-matrix element for the scattering
of a localized colour dipole on a proton, we show that saturation effects can
already be sizable at HERA. The
saturation scale is found to be around 1-1.5 $\mbox{GeV}^2$ for the
highest available energies and for central collisions, and is a
decreasing function of the impact parameter.
}

Presently, there are basically two most discussed theoretical descriptions of the
HERA data for the proton structure functions,
equally successful at small-$x$ but contradictory in their
physical implications. On one hand,
models based on the DGLAP equation fit the data very well.
On the other hand, the recent saturation model\cite{GBW}
reproduces all small-$x$ data with a small number of parameters. The latter assumes a
unitarized cross section but no scale evolution is taken into
account. Its success could mean 
that saturation effects are already important at HERA, at
a scale around $1\ \mbox{GeV}^2$.
However, the very question whether we have reached or not the regime of high
densities at HERA in the perturbative domain has so far not been addressed in
a direct way. We propose here a simple method to 
determine reliably how dense the proton looks.

\section{A brief introduction to saturation}

The HERA physics in the kinematical range considered here is the
physics of photon-proton interactions.
We will denote $W$ the $\gamma^* p$ center-of-mass
energy, $Q^2$ the virtuality of the photon and $x\simeq Q^2/W^2$ the
Bjorken variable.
In a Bjorken frame in which the proton has a large light-cone momentum
$(p^+,0,{\mathbf 0})$ and the photon the momentum
$(0,q^-,{\mathbf q^\bot})$, 
the photon-proton
interaction can be interpreted as the photon probing the partonic
content of the proton. This probe is sensitive to collinear
fluctuations down to transverse distances of
order $1/q^\bot\propto 1/Q$,
and resolves fluctuations occuring over times as short as $1/q^-\propto p^+/W^2$.
Thus by increasing $Q^2$, one reveals more collinear partons.
By raising the center of mass energy $W$, one becomes sensitive to
softer fluctuations.

If one keeps increasing $W$ at fixed $Q$, the density of
partons of size $1/Q$ will eventually be so large that their
wave functions will overlap, and recombination effects will
become sizable. 
The borderline between these two regimes is usually parametrized by
the function $Q^2=Q_s^2(W^2)$. $Q_s$ is called the ``saturation scale''.
The question is whether we have already crossed this line at HERA.

Generally speaking, wave diffraction allows to obtain a picture of a
microscopic object. A mere Fourier transform then relates
the diffractive pattern to the density profile of the object.
This can be done at HERA using diffractive processes, as will
be discussed in the following.

\section{Transparency of the proton to dipoles from diffraction}

At high energy, one can relate diffractive vector meson electroproduction to
dipole-proton elastic scattering. This is the celebrated dipole model\cite{DIPOLENIK},
that we briefly recall here.

In the proton rest frame,
the scattering proceeds in three steps. The virtual photon first breaks up into
a $q\bar q$ pair of size ${\bf r}$, which then scatters off 
the proton, before being recombined
into a meson.
High $W$ guarantees that these processes
are well separated in time. At HERA, the $q\bar q$ pair travels over distances
of a few $10^2$ fermis before entering the interaction region, of
size 1 fm. The recombination takes a comparable ``time''.
This remark leads to the following factorization formula for the
amplitude $\mathcal A$ of this process:
\begin{equation}
\mathcal A(Q^2,W^2,\Delta)=\int d^2{\mathbf r}\,A(W^2,{\mathbf r},\Delta) \int dz\,\psi_Q(z,{\mathbf
r})\psi^*_V(z,{\mathbf r})\ ,
\label{form1}
\end{equation}
where $A$ is the dipole-proton elastic amplitude, $\psi_Q$ and $\psi_V$ are the photon and vector meson
wave functions respectively. The latter has to be modelled: we
considered several models available in the litterature. 
$\Delta$ is the 2-momentum transfer: it is conjugated to the
impact parameter, and this will be the variable with respect to which
we will take the Fourier transform. We define $t\equiv|\Delta|^2$.

Thus the proton sees a beem of asymptotic $q\bar q$ dipoles,
distributed according to the overlap of the photon and meson wave
functions (second integral in formula~(\ref{form1})).
On very general grounds, one can argue that this function is sharply peaked
around a value $r_Q$ of the dipole size scaling like
$1/\sqrt{Q^2+M_V^2}$ ($M_V$ is the mass of the meson, see \cite{DIPOLENIK}).
Hence the amplitude can be approximated by 
$\mathcal A(Q^2,W^2,\Delta)\simeq A(W^2,r_Q,\Delta)\times N(Q)$
where $N(Q)$ is the normalization of the dipole distribution. Note that
$N(Q)$ is {\it a priori} model-dependent, but in practice
this dependence is very weak.

Once we have extracted the dipole-proton amplitude from the HERA data
in this way, we perform a Fourier analysis, by defining the
function
\begin{equation}
S(W^2,r_Q,b)=1-\frac12\int\frac{d^2\Delta}{(2\pi)^2}\, e^{-i{\bf
b}\Delta} A(W^2,r_Q,\Delta)\ .
\end{equation}
One easily checks that $S(W^2,r,b)$ is the $S$-matrix
element for the scattering of a
localized dipole of given size $r$ at given impact parameter $b$.
This holds in particular because at high energy such a dipole state
can only be absorbed: the $S$-matrix is diagonal and its eigenvalues
are real between $0$ and $1$. They can be interpreted as kind of
``transparency coefficients''.
Furthermore, note that $1\!-\!S^2(W^2,r,b)$ is the probability for such a
dipole state to undergo inelastic scattering.

\section{A picture of the proton at HERA and its interpretation}

\begin{figure}[t]
\epsfxsize=22.8pc 
\centerline{\epsfbox{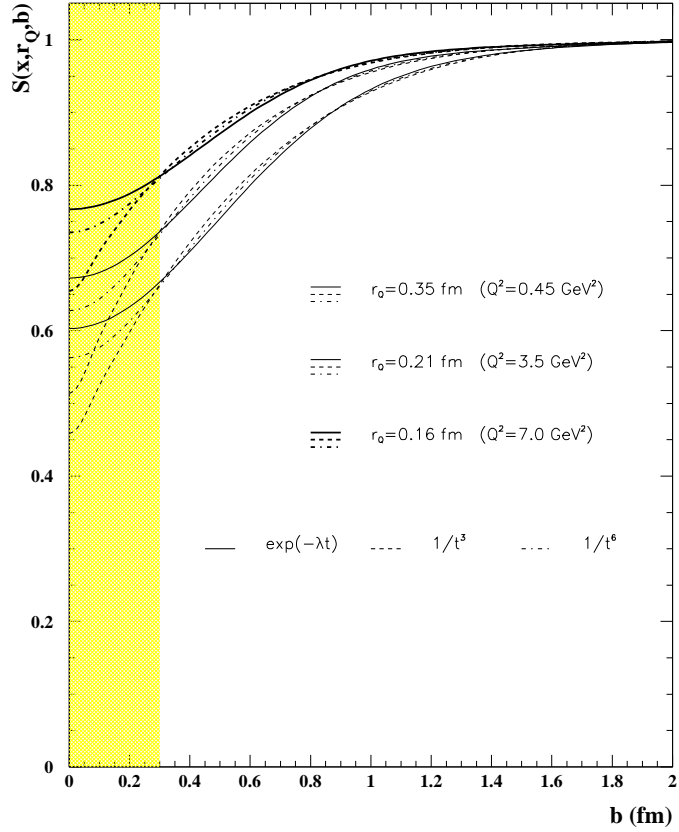}} 
\caption{$S$-matrix element as a function of $b$, for $x=5\cdot 10^{-4}$.  Three values of
$Q^2$ are considered, and three different extrapolations of the data at large
momentum transfer ($\exp(-\lambda t)$, $t^{-3}$, $t^{-6}$).
The sensitivity to this
extrapolation is high in the grey zone.
\label{fig}}
\end{figure}

We have applied the procedure described above to the HERA data.
We took the available data for diffractive production of longitudinal
$\rho$ mesons from the H1 collaboration\cite{H1}. 
$x$ was taken small, around $5\cdot 10^{-4}$. Three different
values of $Q^2$ were considered, $Q^2=7,\,3.5$, and $0.45\
\mbox{GeV}^2$, corresponding to dipole sizes of about $0.16,\,
0.21$ and $0.35\ \mbox{fm}$ respectively. The data was
available only for $t<0.6\ \mbox{GeV}^2$,
so we had to extrapolate them to larger values of $\Delta$ before performing the
Fourier transform. We made three such assumptions. The results are
represented on fig.~\ref{fig}. We get a reliable
extraction of $S(b)$ down to $b=0.3\ \mbox{fm}$.

We see that for say $Q^2<2\ \mbox{GeV}^2$ and $b<0.4\ \mbox{fm}$,
the probability of inelastic interaction is more than 50\%.
Hence the proton looks already quite black in its central region at
present HERA energies.
To see whether this ``blackening'' occurs in the perturbative regime, we
estimate the saturation scale. For this purpose, it is necessary to choose a
parametrisation for $S$. Let us consider the
following Glauber-like form: 
\begin{equation}
S(W^2,r,b)=e^{-Q_s^2(W^2,b)r^2/4}\ ,
\end{equation}
where the fact that $Q_s$ controls the onset of the
saturation regime is quite explicit.
From this formula we now extract the saturation scale and obtain 
$Q_s^2\simeq 1-1.5\ \mbox{GeV}^2$ for $b=0.3\ \mbox{fm}$
and $Q_s^2\simeq 0.2\ \mbox{GeV}^2$ for $b=1.0\ \mbox{fm}$.

We refer the reader to the paper \cite{ref} for all the details of
this analysis and for more results.

\section*{Acknowledgments}

The results presented here were obtained in collaboration with
Dr A.M. Sta\'sto and Pr A.H. Mueller. I thank Dr J.-Y. Ollitrault for
a critical reading of the manuscript.
I also wish to acknowledge support from the EU Framework TMR program, contract FMRX-CT98-0194.

\bibliography{mesons}

\end{document}